\documentclass[aps,prl,twocolumn,showpacs,preprintnumbers,amsmath,superscriptaddress,amssymb,floats,nofootinbib]{revtex4} 
\usepackage{graphicx}
\usepackage[usenames]{color}
\usepackage{ulem}
\begin{document}

\def\Journal#1#2#3#4{{#1} {\bf #2}, #3 (#4)}
\def\NCA{\rm Nuovo Cimento}
\def\NPA{{\rm Nucl. Phys.} A}
\def\NIM{\rm Nucl. Instr. Meth.}
\def\NIMA{{\rm Nucl. Instr. and Meth.} A}
\def\NPB{{\rm Nucl. Phys.} B}
\def\PLB{{\rm Phys. Lett.}  B}
\def\PRL{\rm Phys. Rev. Lett.}
\def\PRD{{\rm Phys. Rev.} D}
\def\PRC{{\rm Phys. Rev.} C}
\def\PR{\rm Phys. Rev.}
\def\ZPC{{\rm Z. Phys.} C}
\def\JPG{{\rm J. Phys.} G}


\title{A measurement of the differential cross section for the reaction $\gamma n \rightarrow \pi^{-} p$ from deuterium}

\newcommand*{\DUKE}{Duke University, Durham, North Carolina 27708}
\affiliation{\DUKE}
\newcommand*{\OHIOU}{Ohio University, Athens, Ohio  45701}
\affiliation{\OHIOU}
\newcommand*{\MSSTATE}{Mississippi State University, Mississippi State, MS 39762}
\affiliation{\MSSTATE}
\newcommand*{\JLAB}{Thomas Jefferson National Accelerator Facility, Newport News, Virginia 23606}
\affiliation{\JLAB}
\newcommand*{\INFNFR}{INFN, Laboratori Nazionali di Frascati, 00044 Frascati, Italy}
\affiliation{\INFNFR}
\newcommand*{\GWU}{The George Washington University, Washington, DC 20052}
\affiliation{\GWU}
\newcommand*{\SACLAY}{CEA-Saclay, Service de Physique Nucl\'eaire, 91191 Gif-sur-Yvette, France}
\affiliation{\SACLAY}


\newcommand*{\ASU}{Arizona State University, Tempe, Arizona 85287-1504}
\newcommand*{\ASUindex}{1}
\affiliation{\ASU}
\newcommand*{\CSU}{California State University, Dominguez Hills, Carson, CA 90747}
\newcommand*{\CSUindex}{2}
\affiliation{\CSU}
\newcommand*{\CMU}{Carnegie Mellon University, Pittsburgh, Pennsylvania 15213}
\newcommand*{\CMUindex}{3}
\affiliation{\CMU}
\newcommand*{\CUA}{Catholic University of America, Washington, D.C. 20064}
\newcommand*{\CUAindex}{4}
\affiliation{\CUA}
\newcommand*{\CNU}{Christopher Newport University, Newport News, Virginia 23606}
\newcommand*{\CNUindex}{6}
\affiliation{\CNU}
\newcommand*{\UCONN}{University of Connecticut, Storrs, Connecticut 06269}
\newcommand*{\UCONNindex}{7}
\affiliation{\UCONN}
\newcommand*{\ECOSSEE}{Edinburgh University, Edinburgh EH9 3JZ, United Kingdom}
\newcommand*{\ECOSSEEindex}{8}
\affiliation{\ECOSSEE}
\newcommand*{\FU}{Fairfield University, Fairfield CT 06824}
\newcommand*{\FUindex}{9}
\affiliation{\FU}
\newcommand*{\FIU}{Florida International University, Miami, Florida 33199}
\newcommand*{\FIUindex}{10}
\affiliation{\FIU}
\newcommand*{\FSU}{Florida State University, Tallahassee, Florida 32306}
\newcommand*{\FSUindex}{11}
\affiliation{\FSU}
\newcommand*{\ECOSSEG}{University of Glasgow, Glasgow G12 8QQ, United Kingdom}
\newcommand*{\ECOSSEGindex}{13}
\affiliation{\ECOSSEG}
\newcommand*{\ISU}{Idaho State University, Pocatello, Idaho 83209}
\newcommand*{\ISUindex}{14}
\affiliation{\ISU}
\newcommand*{\INFNGE}{INFN, Sezione di Genova, 16146 Genova, Italy}
\newcommand*{\INFNGEindex}{16}
\affiliation{\INFNGE}
\newcommand*{\ORSAY}{Institut de Physique Nucleaire ORSAY, Orsay, France}
\newcommand*{\ORSAYindex}{17}
\affiliation{\ORSAY}
\newcommand*{\ITEP}{Institute of Theoretical and Experimental Physics, Moscow, 117259, Russia}
\newcommand*{\ITEPindex}{18}
\affiliation{\ITEP}
\newcommand*{\JMU}{James Madison University, Harrisonburg, Virginia 22807}
\newcommand*{\JMUindex}{19}
\affiliation{\JMU}
\newcommand*{\KYUNGPOOK}{Kyungpook National University, Daegu 702-701, Republic of Korea}
\newcommand*{\KYUNGPOOKindex}{20}
\affiliation{\KYUNGPOOK}
\newcommand*{\UNH}{University of New Hampshire, Durham, New Hampshire 03824-3568}
\newcommand*{\UNHindex}{21}
\affiliation{\UNH}
\newcommand*{\NSU}{Norfolk State University, Norfolk, Virginia 23504}
\newcommand*{\NSUindex}{22}
\affiliation{\NSU}
\newcommand*{\ODU}{Old Dominion University, Norfolk, Virginia 23529}
\newcommand*{\ODUindex}{24}
\affiliation{\ODU}

\newcommand*{\UBP}{Universit\'e Blaise Pascal, Laboratoire de Physique Corpusculaire CNRS/IN2P3 F-63177 Aubi\`ere France}
\affiliation{\UBP}

\newcommand*{\RPI}{Rensselaer Polytechnic Institute, Troy, New York 12180-3590}
\newcommand*{\RPIindex}{25}
\affiliation{\RPI}
\newcommand*{\URICH}{University of Richmond, Richmond, Virginia 23173}
\newcommand*{\URICHindex}{26}
\affiliation{\URICH}
\newcommand*{\MOSCOW}{Skobeltsyn Nuclear Physics Institute, Skobeltsyn Nuclear Physics Institute, 119899 Moscow, Russia}
\newcommand*{\MOSCOWindex}{27}
\affiliation{\MOSCOW}
\newcommand*{\SCAROLINA}{University of South Carolina, Columbia, South Carolina 29208}
\newcommand*{\SCAROLINAindex}{28}
\affiliation{\SCAROLINA}
\newcommand*{\UNIONC}{Union College, Schenectady, NY 12308}
\newcommand*{\UNIONCindex}{30}
\affiliation{\UNIONC}
\newcommand*{\UTFSM}{Universidad T\'{e}cnica Federico Santa Mar\'{i}a, Casilla 110-V Valpara\'{i}so, Chile}
\newcommand*{\UTFSMindex}{31}
\affiliation{\UTFSM}
\newcommand*{\VIRGINIA}{University of Virginia, Charlottesville, Virginia 22901}
\newcommand*{\VIRGINIAindex}{32}
\affiliation{\VIRGINIA}
\newcommand*{\WM}{College of William and Mary, Williamsburg, Virginia 23187-8795}
\newcommand*{\WMindex}{33}
\affiliation{\WM}
\newcommand*{\YEREVAN}{Yerevan Physics Institute, 375036 Yerevan, Armenia}
\newcommand*{\YEREVANindex}{34}
\affiliation{\YEREVAN}

\newcommand*{\NOWVIRGINIA}{University of Virginia, Charlottesville, Virginia 22901}
\newcommand*{\NOWJLAB}{Thomas Jefferson National Accelerator Facility, Newport News, Virginia 23606}
\newcommand*{\NOWGWU}{The George Washington University, Washington, DC 20052}
\newcommand*{\NOWECOSSEE}{Edinburgh University, Edinburgh EH9 3JZ, United Kingdom}

\author {W.~Chen}
\affiliation{\DUKE}
\author {T.~Mibe}
\affiliation{\OHIOU}
\author {D.~Dutta}
\affiliation{\MSSTATE}
\author {H.~Gao}
\affiliation{\DUKE}
\author {J.M.~Laget}
\affiliation{\SACLAY}
\affiliation{\JLAB}
\author {M.~Mirazita}
\affiliation{\INFNFR}
\author {P.~Rossi}
\affiliation{\INFNFR}
\author {S.~Stepanyan}
\affiliation{\JLAB}
\author {I.I.~Strakovsky}
\affiliation{\GWU}


\author {M.J.~Amaryan} 
\affiliation{\ODU}
\affiliation{\YEREVAN}
% ...

\collaboration{The CLAS Collaboration}
\noaffiliation

\date{\today}

\begin{abstract}
 We report a measurement of the differential cross section for the 
$\gamma n \rightarrow \pi^- p$ process from the CLAS detector at 
Jefferson Lab in Hall B for photon energies between 1.0 and 3.5 GeV and pion 
center-of-mass (c.m.) angles ($\theta_{c.m.}$) between 50$^\circ$ and 115$^\circ$. 
We confirm a previous indication of a broad enhancement 
around a c.m.~energy ($\sqrt{s}$) of 2.2 GeV at $\theta_{c.m.}=90^\circ$ 
in the scaled differential cross section, 
$s^7 {\frac{d\sigma}{dt}}$. 
Our data show the angular dependence of this enhancement as the scaling region is approached in the kinematic region from 
70$^\circ$ to 105$^\circ$. 

\end{abstract}

\maketitle

The $\gamma n \rightarrow \pi^- p$, $\gamma p \rightarrow \pi^+ n$
and $\gamma p \rightarrow \pi^0 p$ reactions are essential probes
of the transition from meson-nucleon
degrees of freedom to quark-gluon degrees of freedom in exclusive processes.
The Constituent Counting Rule (CCR)~\cite{farrar,brodsky} was proposed as a signature for the
search of such a transition. According to CCR, the differential cross section for high
energy exclusive two-body reactions at a fixed c.m.~angle scales as
$d\sigma/dt \propto s^{-(n-2)}$. Here $n$ is the total number of point-like 
particles and gauge fields in the initial plus final states and $s$ and $t$ 
are the invariant
Mandelstam variables for the total energy squared and the four-momentum
transfer squared, respectively.
In the last decade or so, an all-orders demonstration
of counting rules for hard exclusive processes has been shown arising
from correspondence between the anti-de Sitter space and the conformal 
field theory~\cite{cft}, which connects superstring theory to QCD.

The differential cross section for many exclusive 
reactions~\cite{anderson76,white94},
at high energy and large momentum transfer, appears to obey the CCR, and
in recent years, the scaling behavior has been observed also
in deuteron photo-disintegration~\cite{slac,bochna,elaine,clas1,rossi} at
a surprisingly low transverse momentum value above about 1.1 GeV/c~\cite{elaine,rossi}.
In addition to the early onset of scaling, some exclusive processes
such as $p p$~\cite{ppdata,hendry} and $\pi p$~\cite{hendry,pidata} elastic scattering, 
show a striking oscillation in
the scaled differential cross section about the predicted
quark counting rule behavior.

The CCR scaling behavior was studied in $\pi^0$~\cite{jenkins} 
and $\pi^+$ photoproduction from the proton
~\cite{jenkins,zhu,e94104_longpaper}, and 
in $\pi^-$ for the first time in the Thomas Jefferson National Accelerator Facility 
(Jefferson Lab) Hall A experiment E94-104~\cite{zhu,e94104_longpaper} 
using a deuterium target and an untagged bremsstrahlung photon beam.
The data of the $ \gamma n \rightarrow \pi^- p$ process 
exhibit an overall CCR scaling behavior at 
$\theta_{c.m.}=70^{\circ}$ and $90^{\circ}$, 
similar to what was observed in the $\pi^{+}$ 
channel at similar c.m.~angles. The data~\cite{zhu} from both the 
 $ \gamma n \rightarrow \pi^- p$ and the 
 $ \gamma p \rightarrow \pi^+ n$
processes at $\theta_{c.m.}=90^{\circ}$ seem to hint at some oscillatory scaling 
behavior. Such oscillatory scaling behavior could be explained as suggested recently by Refs.~\cite{zhao,jxd,dutta-gao}.
 The data also suggest that a transverse momentum
of around 1.2 GeV/c might be the scale governing the onset of scaling, 
consistent with what has been observed in deuteron 
photodisintegration~\cite{elaine,rossi}.
One very interesting feature of the data is an apparent enhancement in the 
scaled differential cross section at $\theta_{c.m.}=90^\circ$ and
at $\sqrt{s}$ range approximately 
from 1.8 GeV to 2.5 GeV.
Furthermore, the scaled differential cross section 
drops by a factor of about 4 in a very narrow c.m.~energy region (few hundreds of MeV) around 2.5 GeV.

The sudden drop in the scaled differential cross section may shed light on the
transition between the aforementioned physical pictures.
It is important to understand the nature of the enhancement followed by the
dramatic drop in the scaled cross section and to test
the onset of scaling behavior in pion photoproduction. This
requires a detailed investigation of the pion photoproduction cross 
section in the $\sqrt{s}$ range from 1.8 to 2.5 GeV with very fine photon 
energy bins. (However, this energy range would not allow for a confirmation 
or refutation of the oscillatory scaling behavior hinted by experiment 
E94-104~\cite{zhu}.) 
In this paper, we report 
such a detailed study using high statistics data from the 
Jefferson Lab CEBAF Large Acceptance Spectrometer (CLAS)~\cite{CNIM} 
in Hall B taken during the g10 running period~\cite{g10}.

The CLAS instrumentation was designed 
to provide large coverage of charged particles ($8^{\circ}\le\theta\le140^{\circ}$). 
It is divided into six sectors by six superconducting 
coils which generate a toroidal magnetic field. Each sector acts as an independent detection system 
that includes drift chambers (DC), Cerenkov counters (CC), 
scintillation counters (SC) 
and electromagnetic calorimeters (EC). The drift 
chambers determine the trajectories of charged particles. With the 
magnetic field generated by the superconducting coils, 
the momenta of the charged particle can be determined from the 
curvature of the trajectories. 
The scintillation counters measure the 
time-of-flight and provide charged particle identification when combined
with the momentum information from the drift chambers. 
Details about the CLAS can be found in Ref.~\cite{CNIM}.

A 24-cm long liquid-deuterium target 
was employed with the target cell 
positioned 25 cm upstream from the CLAS nominal center. A tagged-photon beam~\cite{tagger}
generated by a 3.8-GeV electron beam incident on a gold radiator with a radiation length of $10^{-4}$,
corresponded to a maximum $\sqrt{s}$ 
of 2.8 GeV for the process of interest.
The event trigger required at least two charged particles in different 
sectors. 
Two magnetic field settings were used during the experiment, corresponding 
to a low-field setting 
(with toroidal magnet current I=2250 A) for better forward angle coverage 
, and a high-field setting (I=3375 A) for better momentum resolution.
About $10^{10}$ triggers were collected during the g10 running period of about 
two months.

\begin{figure}
\begin{minipage}[t]{0.23\textwidth}
\includegraphics[width=\textwidth]{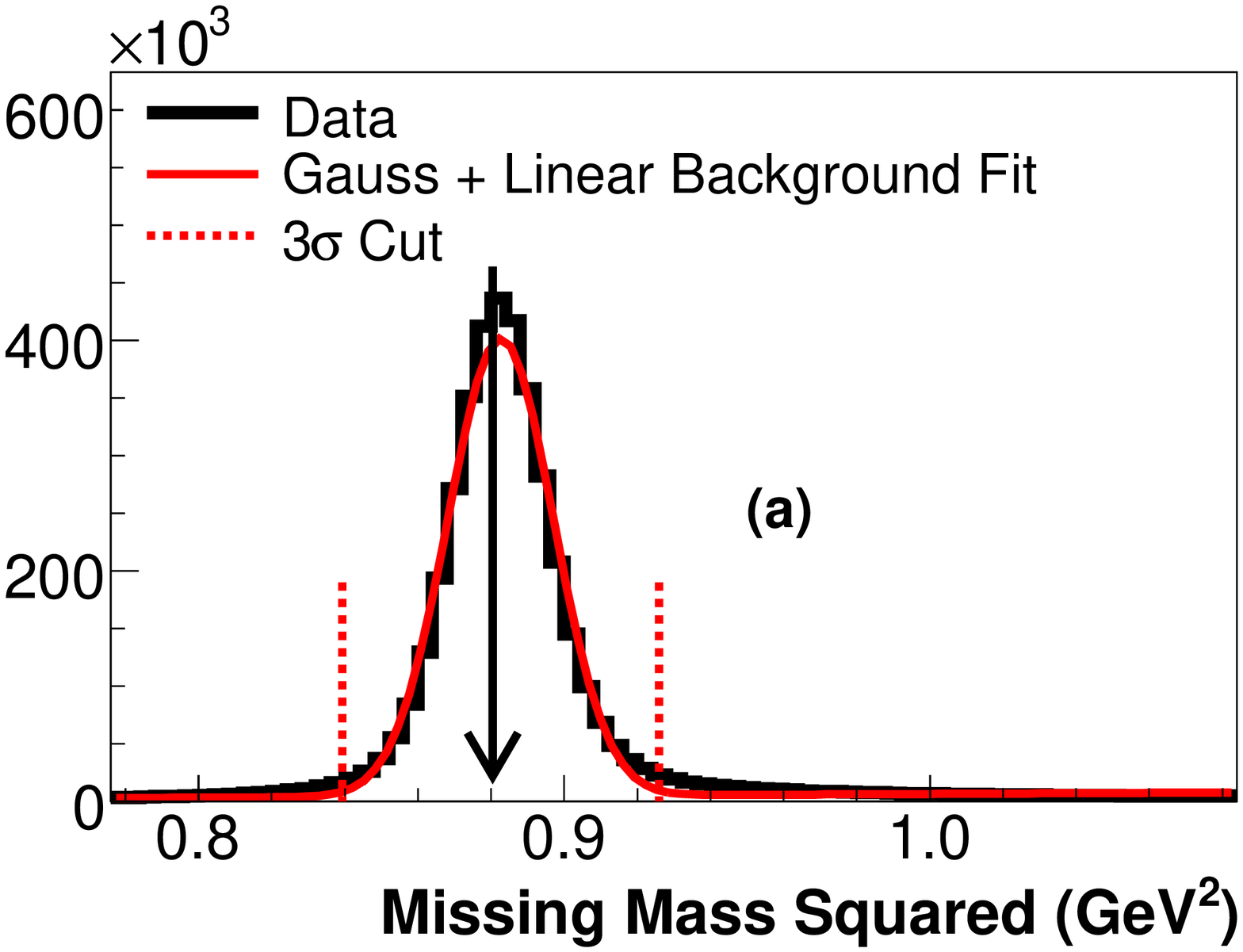}
\end{minipage}
\hfill
\begin{minipage}[t]{0.23\textwidth}
\includegraphics[width=\textwidth]{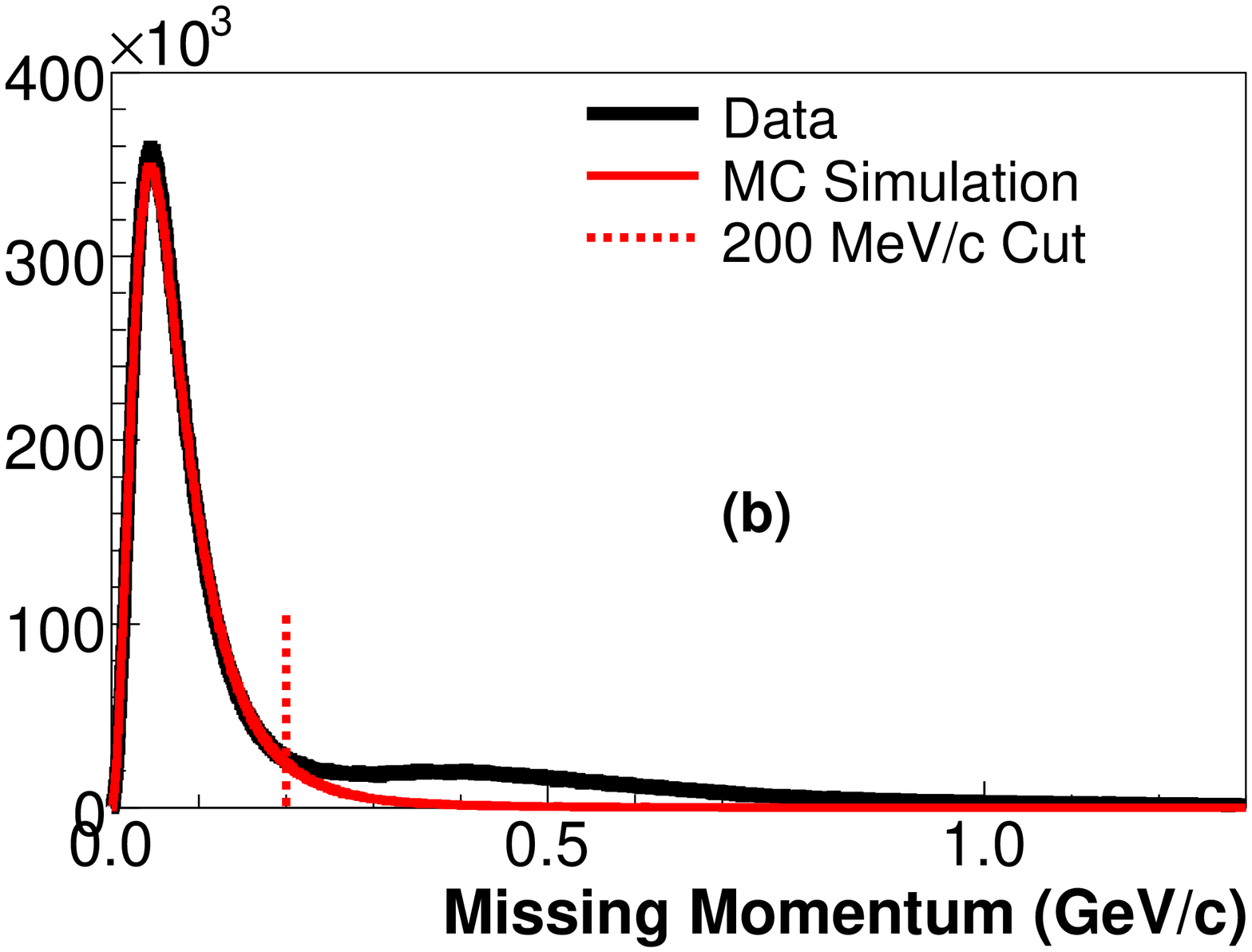}
\end{minipage}
\caption[]{ (color online). 
(a): Reconstructed missing mass squared of the spectator proton 
fitted with a Gaussian plus linear function. 
The arrow indicates the mass squared of the proton;
(b): Reconstructed spectator proton momentum (missing momentum) from
this experiment together with a Monte Carlo simulation.}
\label{fig:missingmomentum}
\end{figure}

The raw data collected from the experiment were first processed
to calibrate and 
convert the information from the detector subsystems to physical variables for detected particles 
such as energy, momentum, position and timing information. 
The events of interest for which the photon coupled to the neutron 
inside the deuteron, were selected by ensuring a proton and a $\pi^{-}$ in the
final state. The difference of the reconstructed time of photon and charged particles at the reaction vertex 
was required to be within 1 ns to ensure that they came from the same 
accelerator electron bunch, which had a period of 2.004 ns.
The momentum of the spectator proton in the deuteron is mostly below 200 MeV/c 
and is therefore not detected by CLAS.
The 4-momentum of the undetected proton was reconstructed by energy-momentum 
conservation. Only events with missing 
mass around the proton mass were selected to make sure that
the missing particle was the 
undetected proton. Shown in Fig.~\ref{fig:missingmomentum}(a) 
is a typical reconstructed missing mass squared distribution.
A $3\sigma$ cut was applied to identify the proton. 
Monte Carlo simulations for the $\gamma n \rightarrow \pi^- p$ process 
based on a phase space generator
have been carried out to determine the acceptance. 
In the simulation, the neutron momentum distribution 
inside the deuteron is based on the deuteron wave function obtained from 
the Bonn potential ~\cite{bonn}.
Fig.~\ref{fig:missingmomentum}(b) shows the reconstructed 
proton momentum from the experimental data and the simulation.
The excellent agreement between the data and the 
Monte Carlo for a missing momentum below 200 MeV/c justified the cut we used 
(shown by the dashed line) in our analysis to select the quasifree events 
of $\gamma n \rightarrow \pi^- p$ from deuterons.  

To extract the cross section, the aforementioned phase space based 
simulation is used to correct for events lost
due to geometrical constraints and detector inefficiencies.
The response of the CLAS detector was simulated in GEANT.
More than $10^8$ of events were generated and 
passed through the simulation.  
The simulated data were then processed to incorporate the subsystem 
efficiencies and resolutions extracted from the experiment. 
The DC wire efficiency and SC efficiency were
studied in detail. The ``excluded-layer method''~\cite{dc} 
was used to study the DC wire
efficiency and identify the bad DC regions. The SC efficiency was
extracted by studying the SC occupancies. The correction due to 
the SC inefficiency is about 20\% for the $\gamma n \rightarrow \pi^- p$ 
channel. All the simulated data 
were then processed by the same software used in the real data processing
and analysis. 
The ratio between the events that passed the simulation and 
the generated events is 
a product of the detector efficiency and the acceptance.

The final state interaction (FSI) effects must be taken into account before 
one extracts cross sections on the neutron since a deuteron target is used. 
The FSI correction is estimated according to 
the Glauber formulation~\cite{trans} and this correction is about 20\%.

The differential cross section in the c.m.~frame of the $\gamma n$ system 
is then given by
\begin{equation}
\frac{d\sigma}{d\Omega_{c.m.}}=\frac{N}{t_{G}\epsilon}\frac{1}{N_{\gamma}}\frac{A}{{\rho}LN_{A}}\frac{1}{d\Omega_{c.m.}},
\end{equation}
where $t_{G}$ is the correction~\cite{trans} for the FSI, $\epsilon$ is the product of the detector efficiency and acceptance, $N$ is 
the number of events, $N_{\gamma}$ is the total number of photons 
incident on the target,
and $A$, $N_{A}$, $L$, $\rho$ are deuteron atomic mass, 
Avogadro's number, target length and target density, respectively.
The scaled differential cross section is defined as
\begin{equation}
s^7\frac{d\sigma}{dt}=s^7\frac{d\sigma}{d\Omega_{c.m.}}\frac{d\Omega_{c.m.}}{dt
}\\
=s^7\frac{d\sigma}{d\Omega_{c.m.}}\frac{\pi}{E_{c.m.}^{\gamma}p_{c.m.}^{\pi^{-}}}
,
\end{equation}
where $E_{c.m.}^{\gamma}$ and $p_{c.m.}^{\pi^{-}}$ are the photon energy and
$\pi^-$ momentum in the c.m.~frame, respectively. 
The results from the high magnetic field setting are consistent with
those from the low magnetic field setting within systematic uncertainties.

There are three major sources of systematic uncertainties: the luminosity, the FSI correction, and the background.
We studied the target thickness fluctuations as seen by the beam, as well as the 
run-dependent, and beam-current-dependent 
fluctuations of the normalized yield. All of them contribute to 
the uncertainty in the luminosity, and in total this uncertainty is 
less than 5\%. The uncertainty of the Glauber calculation for the 
FSI correction was estimated to be 5\% in Ref.~\cite{zhu}. To study the model
uncertainty in calculating the FSI correction, we carried out 
another calculation 
using the approach of Ref.~\cite{laget}. Both methods agree within 10\%. 
 A 10\% systematic uncertainty to the differential cross section is 
assigned for the FSI correction.
The background in the missing mass peak region 
is about 2\% - 7\% depending on the photon energy and an example is shown in
 Fig.~\ref{fig:missingmomentum} (left).
According to Monte Carlo simulations, the background 
could come from the poorly reconstructed 
real events due to the DC resolution. Therefore, 
no background was subtracted in this analysis, instead the fitted background
was assigned as the systematic uncertainty. The overall systematic uncertainty 
is between 11\% to 13\% on the extracted differential cross sections.

\begin{figure}[htbp]
{\includegraphics*[height=8.5cm]{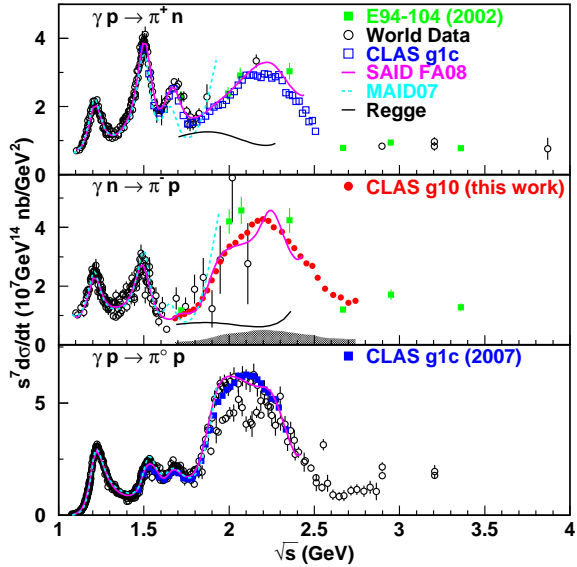}}
\caption[]{(color online). Scaled differential cross section $s^7{\frac{d\sigma}{dt}}$ as a
function of $\sqrt{s}$ for $\theta_{c.m.} = 90^\circ$ for three different channels. The upper panel is for the
$\gamma p \rightarrow \pi^+ n$ process, the middle panel is for the
$\gamma n \rightarrow \pi^- p$ process,
 and the lower panel is for the
$\gamma p \rightarrow \pi^0 p$ process.
The green solid squares are
results from Ref.~\cite{zhu} and
the results from this
experiment are shown as red solid circles. Results from Dugger
{\it et al.}~\cite{fa06}
on neutral pion production are shown as blue solid squares. 
The blue open squares are recent CLAS data 
on $\pi^{+}$ production~\cite{fa08}.
The SAID FA08
results~\cite{fa08} are shown as the magenta curves in all three panels.
The prediction from a Regge approach~\cite{regge} is shown in the top and middle panels by black curves.
The black open circles are the world data collected from Refs.~\cite{anderson76,world_data}.
}
\label{scaling_gn_gp}
\end{figure}

Fig.~\ref{scaling_gn_gp} shows the  
scaled differential cross section $s^7{\frac{d\sigma}{dt}}$ as a 
function of $\sqrt{s}$ for $\theta_{c.m.} = 90^\circ$ for three different channels.
The results from this 
experiment are shown 
in the middle panel as red solid circles with statistical uncertainties, 
 and the systematic uncertainty is shown as a band.
The error bars for E94-104~\cite{zhu} include both 
the statistical and systematic
uncertainties, while only statistical uncertainties are shown for 
the $\pi^0$ data~\cite{fa06} and the $\pi^{+}$ data~\cite{fa08}.
All other world data are collected from Refs.~\cite{anderson76,world_data}.
There are three distinct features shown in the data:
a broad enhancement around $\sqrt{s}$ of 2.1 GeV; a 
marked fall-off of the differential cross section in a narrow energy window of
about 300 MeV above this enhancement; and the suggested~\cite{zhu} onset of the CCR scaling for $\sqrt{s}$ around 2.8 GeV.
The second feature was suggested by Jefferson Lab experiment E94-104~\cite{zhu} 
(shown as green solid squares) and the 
$\pi^- p$ total scattering cross section data~\cite{pip}. 
The drastic fall-off of the cross section has now been firmly 
established by the results from this experiment.
Also shown are the results of the SAID FA08 partial wave analysis~\cite{fa08} (magenta), the
MAID07 model~\cite{maid} (blue), and the prediction from a Regge 
approach~\cite{regge} (black).

The Regge approach does not describe our data and 
the deviation is speculated to be due to baryon resonances~\cite{regge}.
While the SAID FA08 fit has been greatly improved by the CLAS 
$\pi^0$~\cite{fa06} and the $\pi^+$ data~\cite{fa08}, 
it does not give as good a description of the data near the peak of the 
enhancement. Further, it lacks the constraint 
on the $\pi^{-}$ channel and 
does not describe our data well above 2.4 GeV in $\sqrt{s}$.
The precision data presented here will help to further constrain the 
SAID fit and will allow for a determination of the corresponding neutron 
electromagnetic parameters for 4-star PDG resonances. 
These studies will be reported 
in a future publication.

\begin{figure}[t]
{\includegraphics*[height=8.5cm]{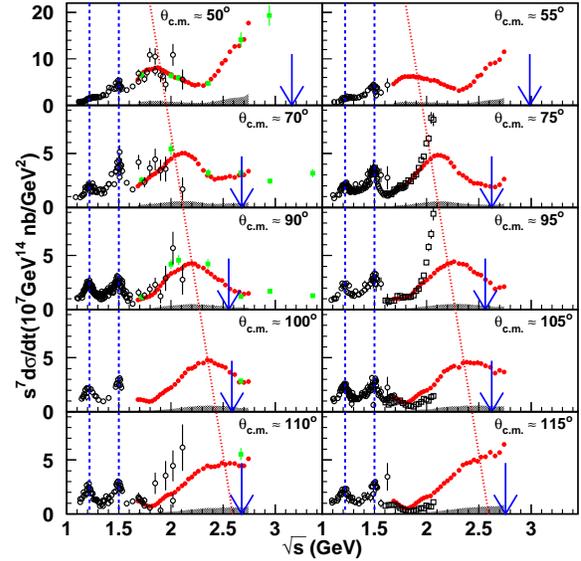}}
\caption[]{(color online). Scaled differential cross section $s^7{\frac{d\sigma}{dt}}$ as a 
function of $\sqrt{s}$ for $\theta_{c.m.} = 50^\circ$ to $115^\circ$. The arrows 
indicate the location of $\sqrt{s}$ corresponding to a transverse momentum 
value of 1.1 GeV/c. The green solid squares are results from Ref.~\cite{zhu}.
The results from this experiment are shown as 
red solid circles. The black open circles and open squares 
are the world data collected from Refs.~\cite{anderson76,world_data} and~\cite{besch}, respectively.
Errors on the data from CLAS are the quadratic sums of the statistical 
and systematic uncertainties. The blue dashed lines indicate the known resonances, and the red dotted lines illustrate the angular dependent feature of the broad enhancement structure discussed in the text.} 
\label{scaling_plot4}
\end{figure}

Fig.~\ref{scaling_plot4} shows the 
scaled differential cross section $s^7{\frac{d\sigma}{dt}}$ as a 
function of $\sqrt{s}$ for $\theta_{c.m.} = 50^\circ$ to $115^\circ$ with an angular bin size of $5^\circ$ for the
$\gamma n \rightarrow \pi^- p$ process. As in Fig.~2, the systematic 
uncertainties are shown as bands in Fig.~3.
The blue arrows 
indicate the location of $\sqrt{s}$ corresponding to a pion transverse momentum 
($p_T$) of 1.1 GeV/c. This $p_T$ value was suggested to govern the scaling onset by Refs.~\cite{elaine,rossi}.
 We note the large discrepancy between 
our results and those from Ref.~\cite{besch} at $\theta_{c.m.} = 75^\circ$ and 
95$^\circ$. We also note that the SAID fits~\cite{fa06,fa08} did not 
include data from Ref.~\cite{besch}. 
An angular-dependent feature in the scaled differential cross section 
is clearly seen in our data. The aforementioned broad enhancement 
around a 
$\sqrt{s}$ value of 2.1 GeV at $\theta_{c.m.}= 90^\circ$  
seems to shift
as a function of $\theta_{c.m.}$ from $\sqrt{s}$ of 1.80 GeV at $50^\circ$ 
to 2.45 GeV at $105^\circ$ as shown by the red dotted lines.
Our studies show that such behavior is not an artifact of the 
$s^7$ scaling factor.
It is not clear whether this enhancement dies off for 
$\theta_{c.m.} > 105^\circ$ or whether it shifts to further higher energies.
The blue dotted lines indicate the locations of the nucleon resonances around 
1.2 GeV and 1.5 GeV which, as expected, do not change with $\theta_{c.m.}$.
However, such an angular dependent scaling behavior is not present in
the $\pi^+$ and $\pi^0$ channels from the proton~\cite{note}. 
Our preliminary studies show that such a behavior is not due to the FSI correction, while more complete calculations are in progress.

The approach to the scaling region is seen in Fig.~\ref{scaling_plot4} at 
the highest $p_T$ kinematics, from $\theta_{c.m.} = 70^\circ$ to 105$^\circ$. 
In the forward angle kinematics of
50$^\circ$, higher energies are necessary to reach a $p_T$ value
of 1.1 GeV/c, suggested by the deuteron photodisintegration 
data~\cite{elaine,rossi} as the 
value for the onset of the scaling behavior.
It is very important
to extend this experiment to much higher photon energies,
such as is feasible at 6 GeV at Jefferson Lab currently 
and at 11 GeV at the energy-upgraded Jefferson Lab facility in the future, 
and to carry out similar measurements 
on the $\gamma p \rightarrow \pi^+ n$ and the 
$\gamma p \rightarrow \pi^0 p$ processes, and polarization measurements 
for all three channels.
Such studies will be essential 
in understanding the nature of the observed enhancement, 
the running behavior of 
the enhancement structure in the $\pi^-$ channel, 
and to understand where and how the transition 
from the nucleon-meson to the quark-gluon degrees of QCD takes place.

We acknowledge the outstanding efforts of the staff  
of the Accelerator and Physics 
Divisions at Jefferson Lab who made this experiment possible. 
This work was supported in part by the U.S.~Department of Energy, 
the National Science Foundation, 
the Italian Istituto Nazionale di Fisica Nucleare, the French 
Centre National de la Recherche Scientifique and Commissariat \`a 
l'Energie Atomique, and the Korea Science and Engineering Foundation.
Jefferson Science Associates (JSA) operates the 
Thomas Jefferson National Accelerator Facility for the 
U.S.~Department of Energy under contract DE-AC05-060R23177.

\end{document}